\documentstyle[pre,aps,epsf]{revtex}
\newcommand \beq{\begin{eqnarray}}
\newcommand \eeq{\end{eqnarray}}
\newcommand{\set}[2]{\newcommand{#1}{#2}}
\set{\pa}{\partial \over \partial\, }
\set{\leftvector}{\stackrel{\leftarrow}{\partial }}
\set{\rightvector}{\stackrel{\rightarrow}{\partial }}
\begin{document}
%\twocolumn[\hsize\textwidth\columnwidth\hsize
%           \csname @twocolumnfalse\endcsname
%\draft
\title{
Debye-Onsager-Relaxation-Effect beyond linear Response and
Antiscreening in Plasma
Systems}
\author{K. Morawetz}
\address{
Fachbereich Physik der Universit\"at Rostock\\
18051 Rostock, Germany
}
\maketitle
\begin{abstract}
The quantum kinetic equation for charged particles in strong electric fields is derived and analyzed with respect to the particle flux. It is found that the applied electric field is screened nonlinearly. The relaxation field is calculated completely quantum mechanically and up to any order in the applied field. The classical limit is given in analytical form. In the range of weak fields the deformation of the screening cloud is responsible for the Debye-Onsager relaxation effect. The result beyond linear response presented here allows to investigate a field regime where no screening cloud is present. The descreening field is determined as a function of thermal energy density of the plasma. For stronger fields the moving charge is accelerated by accumulated opposite charges in front of the particle. This can be understood in analogue to the accoustic Doppler effect. A critical field strength is presented up to which value a thermalized plasma is possible. The range of applicability of the treatment is discussed with respect to applied field strength and space gradients.
\end{abstract}
%\vskip2pc]

\newcommand{\grlo}{\stackrel{>}{<}}
\newcommand{\logr}{\stackrel{<}{>}}

\section{Introduction}

High field transport has become a topic of current interest in various fields of physics. In semiconductors the nonlinear transport effects are accessible due to femtosecond laser pulses and shrink devices \cite{HJ96}. In plasma physics these field effects can be studied within such short pulse periods \cite{THWS96}. One observable of interest is the current or the electrical conductivity which gives access to properties of dense nonideal plasmas \cite{kker86}. In high energy physics the transport in strong electric fields is of interest due to pair creation \cite{KES93}.

In order to describe these field effects one can start conveniently from kinetic theory. Within this approach the crucial question is to derive appropriate kinetic equations which include field effects beyond linear response. The most promising theoretical tool is the Green function technique \cite{HJ96,D84,SL94,SL95}.
The resulting equations show some typical deviations from the ordinary Boltzmann equation: (i) A collision broadening which consists in a smearing out of the elementary energy conservation of scattering. This is necessary to ensure global energy conservation \cite{M94}. (ii) The intracollisional field effect, which gives additional retardation effects in the momentum of the distribution functions. This comes mainly from the gauge invariance.

One of the most important question is the range of applicability of these kinetic equations. Up to which field strengths are such modifications important and appropriate described within one-particle equations ? In \cite{lkaw91} this question has been investigated for semiconductor transport. It was found that for low fields ordinary transport equations are sufficient while for high fields the field effects are absent. The critical value for this range is given by a characteristic time scale of field effects $\tau_F^2=m \hbar /(e E \cdot q)$ compared with the inverse collision frequency. This criterion is a pure quantum one. It remains the question whether there are also criteria in the classical limit. For a plasma system we will find that there is indeed a critical value of the field strength which can be given by classical considerations.

At low fields we expect the linear response regime to be valid. Then the contribution of field effects to the conductivity can be condensed into the Debye- Onsager relaxation effect \cite{k58,KE72,e76,er79,r88,MK92} which was first derived within the theory of electrolytes \cite{DH23,O27,f53,FEK71,KKE66}. Debye has given a limiting law of electrical conductivity \cite{DH23} which stated that the external electric field $E$ on a charge $e$ is diminished in an electrolyte solution by the amount
\beq\label{debye}
\delta E=E \,(1- {\kappa e^2 \over 6 \epsilon_0 T})
\eeq
where $\kappa$ is the inverse screening radius of the screening cloud. This law is interpreted as a deceleration force which is caused by the deformed screening cloud surrounding the charge $e$. Later it has been shown by Onsager that this result has to be corrected to
\beq
E \,(1- {\kappa e^2  \over 3 (2 +\sqrt{2})\epsilon_0 T})
\eeq
if the dynamics of ions is considered.

In this paper we will give the complete result beyond linear response for the static case (\ref{debye})
\beq\label{new}
E \,(1- {\kappa e^2 \over 6 \epsilon_0 T} F(E)).
\eeq
We find that $F(E)$ represents a nonlinear dependence of the Debye- Onsager relaxation- effect on the field strengths (\ref{classical}), which is a result beyond linear response.
We see that the electric field is limited below a value of
\beq\label{cond3}
E<{\kappa T\over e}.
\eeq
Above this limit no quasi- equilibrated transport is possible, i.e. no thermal distributions are pertained in the system. Then we have to take into account nonthermal field dependent distributions which have been employed to study nonlinear conductivity \cite{MSK93,KMSR92}. In particular new types of instabilities are possible \cite{MK93}.

We shall give in this paper a regime below this limit where the surprising effect of antiscreening can occur. We shall see that in the range
\beq
{\kappa T\over e}>E> 0.7 {\kappa T\over e}
\eeq
the sign of $F(E)$ in (\ref{new}) can change which means that already the field blows away the screening cloud. However, the charged particle feels an accumulation of opposite charges in front of its path. This can be considered as an analogue to the Doppler effect. We will call it Charge - Doppler effect (CD- effect).
This has consequences on the conductivity which has to increase within this range. If the field reaches values of $E=0.7 \kappa T/e$ we have no screening cloud.
This effect of descreening has been described by different methods \cite{YKF92,MJ93} .

First we give a short rederivation of the field dependent kinetic equations from the Green's function technique. In the third section we shall derive the field dependent current analytically. This will produce the main result of this paper which is the nonlinear relaxation effect. The complete quantum case is given up to cubic response. The classical result is calculated analytically for complete response up to all orders in the field strength. In the last section we shall discuss some applications of this CD- effect.

\section{Real-Time Green's Function Technique} \label{greens}

We would like to consider a system of charged particles of fermions or
bosons under the
influence of an external constant electric field. Coupling the field in vector
potential gauge we have the Hamiltonian
\begin{eqnarray}\label{Hamilton}
H=\sum_{i} && \left. \int d{\bf r}
\,
\Psi_i^* ({\bf r},t)
\,
H_i(\frac{\hbar}{i}{\bf \nabla_r}-\frac{e}{c} {\bf A}(t))
\,
\Psi_i({\bf r},t) \right. \nonumber\\
+ \frac{1}{2} \sum_{i,j}
&& \left. \int d{\bf \bar{r}} d{\bf r}
\,
\Psi_i^*({\bf r},t)
\,
\Psi_j^*({\bf \bar{r}},t)
\,
V_{i,j}({\bf r}-{\bf \bar{r}})
\,
\Psi_i({\bf r},t)
\,
\Psi_j({\bf \bar{r}},t)
\right. .\nonumber\\
&&
\end{eqnarray}
With the Coulomb potential 
\[
V_{ij}({\bf r}-{\bf \bar{r}})=\frac{e^2Z_iZ_j}{|{\bf r}- \bar{\bf
r}|}.
\]

\subsection{Definitions and Equation of Motion}

In order to describe correlations in highly nonequilibrium
situations, we define
various correlation functions by different products of creation and annihilation
operators

\begin{eqnarray}\label{correlation}
G^> (1,2) & = & <\Psi(1)\Psi^+(2)> \nonumber\\
G^< (1,2) & = & <\Psi^+(2)\Psi(1)>.
\end{eqnarray}
Here $<>$ is the average value with the unknown statistical {\it nonequilibrium}
operator $\rho$ and $1$ denotes the cumulative variables
$({\bf r_1},s_1,t_1...)$ of space, spin, time etc.

From the above definitions we can build a causal function with
the help of the step function $\Theta$

\begin{equation}\label{causal}
G(1,2)=\Theta(t_1-t_2) G^>(1,2)-\Theta(t_2-t_1) G^<(1,2).
\end{equation}

We use the equation of motion
of the creation and annihilation operators to derive
kinetic equations, which may be solved with the appropriate choice of boundary and
initial conditions. Applying the equation of
motion for the field operators in the Heisenberg picture, one
finds the famous
Martin Schwinger equation hierarchy \protect\cite{KB62}, where the one-particle
Green's function $G$ couples to the two-particle one $G_2$, etc.

A formally closed equation can be reached
with the introduction of the self-energy
\begin{equation} \label{selfenergy}
\left. \int d2 V(1-2) G_2(12,1'2^+) \right. =  \left. \int\limits_C d\bar{1}
\Sigma (1,\bar {1}) G(\bar {1},1') \right. .
\end{equation}

In order to obtain solutions and the path of integration $C$, it is
necessary to specify the initial conditions. In many physical
situations,
the condition of weakening of initial correlation is an appropriate choice

\begin{equation}\label{weakening}
\lim_{t_{1} \rightarrow -\infty} G_{2} (121'2') |_{t_{1}=t_{2}+\epsilon}
= G(11')G(22')-G(12')G(21')
\end{equation}
which leads into a vanishing selfenergy in (\ref{selfenergy}).
Therefore we can choose for the path $C$ the ordinary time integration along the real axis if we subtract just terms which ensure this limit. The result reads
\beq \label{way}
&&\left. \int\limits_C d\bar {1} \Sigma (1,\bar {1}) G(\bar {1},1') \right. 
=
\left. \int\limits_{-\infty}^{+\infty} d\bar {1} \left \{
\Sigma(1,\bar {1})
G(\bar {1},1') - \Sigma^< (1,\bar {1}) G^> (\bar {1},1') \right \} \right.
\nonumber\\
&=&\int_{-\infty}^{t_1}\Sigma^> (1,\bar {1}) G^< (\bar {1},1')+
\int_{t_1}^{t_1'}\Sigma^< (1,\bar {1}) G^< (\bar {1},1')+ 
+\int_{t_1'}^{\infty}\Sigma^< (1,\bar {1}) G^> (\bar {1},1')-
\int_{-\infty}^{\infty}\Sigma^< (1,\bar {1}) G^> (\bar {1},1')
\nonumber\\
&&
\eeq
which indeed vanishes for $ t_1 \rightarrow t_1 \rightarrow t_o
=-\infty$.
This can be alternatively formulated as a contour of time integration known as Keldysh contour.
This means with other words that the weakening
of the initial correlation which breaks the time reversal symmetry
is equivalent to the Keldysh contour integration.
With the expressions (\protect\ref{selfenergy})
and (\protect\ref{way}) we can finally write the first equation of the
Martin-Schwinger hierarchy in the form of the Kadanoff-Baym equation
\protect\cite{KB62,Kel64,SL95}
\begin{eqnarray}
  -&i& \left ( G_0^{-1} G^<-G^< G_0^{-1} \right )=
  i \left ( G^R \Sigma^<-\Sigma^< G^A \right ) -i
  \left ( \Sigma^R G^<-G^< \Sigma^A \right )\label{kb}
\end{eqnarray}
where the retarded and advanced functions are introduced as
$A^{R}(1,2)=-i \Theta (t_1-t_2) [A^>\pm A^<]$ and
$A^{A}(1,2)=i \Theta (t_2-t_1) [A^>\pm A^<]$.
Here operator notation is employed where products are understood as integrations
over intermediate variables (time and space) and the upper/lower sign stands for fermions/bosons respectively. The Hartree- Fock drift term
reads
\begin{equation}
  G_0^{-1}(1{1'})=\left (i \hbar {\partial \over \partial t_1}- {({\hbar \over
      i}\nabla_{x_1})^2 \over 2 m} - \Sigma^{HF}(1{1'}) \right ) \delta
  (1-{1'}) \label{hf}
\end{equation}
with the Hartree Fock self energy
\begin{eqnarray} \label{HF}
\Sigma_{HF}(1,1') &=& \left ( \mp \delta (r_1 - r_1') \int dr_2 V(r_1 -r_2)G^<
(r_2t_1'r_2t_1) \right . \nonumber \\
&& \left . + V(r_1 - r_1')G^< (r_1t_1r_1't_1') \right ) \delta (t_1 -t_1')
\end{eqnarray}
where $G(r_2 ,t_1,r_2,t_1)=n(r_2,t_1)$ is the density.

Introducing
Wigner coordinates $T=(t_1+t_2)/2,\tau=t_1-t_2, etc$,
we finally obtain the time diagonal part as
\protect\cite{JW84}

\begin{eqnarray}\label{timediagonal}
\frac{\partial}{\partial T}
&&f_W (p,T) =\int\limits_0^{\infty} d\tau
\left[ \left \{ G^>(p,T-\frac{\tau}{2},-\tau),\Sigma^<(p,T-\frac{\tau}{2},\tau) \right \} \right .-
\left . \left\{ G^<(p,T-\frac{\tau}{2},-\tau),\Sigma^>(p,T-\frac{\tau}{2},\tau) \right \} \right]\nonumber\\
&&
\end{eqnarray}
Here $f_W(p,R,T)=\mp i g^{<}(p,R,T,\tau=0)$ denotes the Wigner distribution function and $\{, \}$ is the anti-commutator.
This equation is exact in time, but according to the assumed slowly varying space dependence
we have used gradient expansion for space variables and dropped all R-dependence for
simplicity. This criterion is discussed in the last section (\ref{space1}).

\subsection{Gauge invariance}

In order to get an unambiguous
way of constructing approximations we have to formulate our theory in gauge invariant way.
This can be done following a procedure known from field theory
\protect\cite{I80}. This method has been applied to high field problems in
\protect\cite{BJ91}.

One can introduce a gauge-invariant Fourier-transform of the difference
coordinates x
\begin{equation}\label{Fouriertrafo}
{\bar g}(k,X) = \int dx \; \;{\rm exp}
\left \{ \; \left . \frac{i}{\hbar} \int\limits_{-\frac{1}{2}}^{\frac{1}{2}}
d\lambda \; x_{\mu}[k^{\mu}+\frac{e}{c}A^{\mu}(X+\lambda
x)] \right .\right \} g(xX).
\end{equation}
For constant electric fields, which will be of interest in the
following, one obtains
for the generalized Fourier-transform
\[
{\bar g}(k,X) = \int dx\;
{\rm e}^{\frac{i}{\hbar}[x_\mu
k^\mu+e{\bf rE}T]} g(x,X),
\]
where the $\chi$ function was chosen in such a way
that the scalar potential is zero $A^\mu =(0,-c {\bf E}T)$.
Therefore, we have the following rule in formulating the kinetic theory
gauge-invariantly
\begin{enumerate}\label{gaugerule}
\item{Fourier transformation of the difference-variable x to canonical momentum p.}
\item{Shifting the momentum  to kinematical momentum
according to $p=k-eET$.}
\item{The gauge invariant functions $\bar {g}$ are given by
\begin{equation}\label{gauge invariant}
g(p,T)=g(k-eET,T)={\bar g}(k,T)={\bar g}(p+eET,T).
\end{equation}
}
\end{enumerate}
We shall make use of these rules in the following sections.
In \cite{Mor94} this procedure is generalized to two- particle Greens functions and leads to the field- dependent Bethe- Salpeter- equation.

\subsection{Spectral function}

The spectral properties of the system are described by the
Dyson equation for the retarded Green function.
A free particle in a uniform electric field,
where the field is represented by a vector potential $E(t)=-\frac 1 c \stackrel{.}{A(T)}$ leads to the following equation
\begin{equation}\label{free}
\left [i\frac{\partial}{\partial t}-\epsilon(p-\frac{e}{c}A(t)) \right ] G^R_0(p,tt')=
\delta (t-t').
\end{equation}
This equation is easily integrated \protect\cite{jau91,kdw87}
\begin{equation}\label{ret}
G^R_0(p,tt')=-i\Theta(t-t')\exp{\;\left [{i \over \hbar}\int\limits_t^{t'}du
\;\epsilon(p-{e \over c}A(u)) \right ]}.
\end{equation}
For free particles and parabolic dispersions,
the gauge invariant spectral function \protect\cite{jau91,kdw87}
follows
\begin{eqnarray}\label{spectrall}
A_0(k,\omega)&=&2\int\limits_0^{\infty}d\tau {\rm cos} \left
( \omega\tau- \frac{k^2}{2m\hbar}\tau-
\frac{e^2E^2}{24m\hbar} \tau^3 \right ) \nonumber \\
&=& \frac{2\pi}{\epsilon_E} Ai \left
(\frac{k^2/2m-\hbar\omega}{\epsilon_E} \right )
\end{eqnarray}
where Ai(x) is the Airy function \protect\cite{a84} and $\epsilon_E=(\hbar^2e^2E^2/8m)^{1/3}$.
It is instructive to verify that (\protect\ref{spectrall}) satisfies the frequency
sum rule.
This interaction-free but field-dependent
retarded Green's function $G_o^R $ can be obtained from the
interaction-free
and field-free
Green's function by a simple Airy transformation \cite{Moa93}. This is an expression of the fact that
the solutions of the Schr\"odinger equation with constant electric
field are Airy-functions.
The retarded functions can therefore be diagonalized
within those eigen-solutions\protect\cite{BKF89,BJ91}. It can be shown that (\ref{spectrall}) remains valid even within a quasiparticle picture \cite{Moa93},
where we have to replace simply the free dispersion $k^2/2m$ by the quasiparticle energy $\epsilon_k$.

\subsection{The Problem of the ansatz}
In order to close the kinetic equation
(\protect\ref{timediagonal}), it is
necessary to know the relation between $G^>$ and $G^<$. This problem is known
as an ansatz and must be constructed consistently with the required approximation
of self-energy.
The conventional way to do this is to change the correlation
functions into the generalized
distribution function and into the spectral one, which is still an exact transformation

\begin{eqnarray}\label{variab}
G^{<}(p\omega RT) &=& A(p \omega RT)F(p \omega RT) \nonumber\\
G^{>}(p\omega RT) &=& A(p \omega RT) (1 \mp F(p \omega RT)),
\end{eqnarray}
with the spectral function
from eq. (\protect\ref{spectrall}).

Assuming the conventional ansatz, we replace the $\omega$ dependence
of the distribution function by their quasiparticle value, which is given by the quasiparticle distribution function $f$

\beq\label{g}
G^<(k\omega RT)&=& A(k\omega RT)\; f(kRT)\nonumber\\
G^>(k\omega RT)&=& A(k\omega RT)\; (1\mp f(kRT)).
\eeq
This is quite good as long as the quasi-particle picture holds and no memory
effects play any role. As we shall see, the
formulation of kinetic equations with high fields is basically connected with
a careful formulation of retardation times. Therefore, the simple
ansatz,
called {\it KB ansatz} fails.

Another obscure discrepancy is the fact that with the old ansatz,
one has some
minor differences in the resulting collision integrals
compared with
the results from the density operator technique. With the old
ansatz, one gets just
one half of all retardation times in the various time arguments \protect\cite{JW84,jau91}.
This annoying
discrepancy remained obscure until the work of Lipavsky, {\it et al.}
\protect\cite{LSV86} where an expression is given
for the $G^<$ function in terms of expansion after various times.
We can write in Wigner coordinates
\begin{eqnarray}\label{gk}
G^<(p,T,\tau) &=&f(p,T-\frac{|\tau|}{2})A(p,\tau,T).
\end{eqnarray}
The generalized- Kadanoff- Baym (GKB) - ansatz is an exact relation as long as the selfenergy is taken in Hartree- Fock approximation.
Together with the requirement of gauge invariance (\protect\ref{gaugerule})
the GKB ansatz finally reads
\begin{eqnarray}\label{Lipavskyg}
G^<(k,T,\tau)&=&f(k-\frac{eE|\tau|}{2},T-\frac{|\tau|}{2})
\,A(k,\tau,T) \nonumber\\
G^>(k,T,\tau)&=&(1\mp f(k-\frac{eE|\tau|}{2},T-\frac{|\tau|}{2}))\,A(k,\tau,T).\nonumber\\
&&
\end{eqnarray}
At the end, we can use the spectral function derived in the preceding
section (\protect\ref{spectrall}) to obtain the resulting ansatz valid for high
applied electric fields

\begin{equation}\label{Lipavsky}
G^<(k\tau RT)=e^{-\frac{i}{\hbar} \left(\epsilon_k \tau + \frac{e^2E^2}{24m}\tau^3\right)}
f \left( k-\frac{eE|\tau|}{2},R,T-\frac{|\tau|}{2} \right).
\end{equation}
In order to get more physical insight into this ansatz one
transforms into the frequency representation
\beq
G^<(k\omega RT)= &2&\int\limits_0^{\infty}d\tau {\rm cos}
\left
(  \omega\tau- \epsilon(k,R,T){\tau \over \hbar}-
\frac{e^2E^2}{24m\hbar} \tau^3 \right )
f(k-\frac{e E\tau}{2},T-\frac{\tau}{2}).
\eeq
Neglecting the retardation in $f$ one recovers the ordinary ansatz
(\ref{g}) with the spectral function (\ref{spectrall}).
The generalized ansatz takes into account
history by an additional memory. This ansatz is superior to the Kadanoff-Baym ansatz in the
case of high external fields in several respects \cite{MJ93}: (i) it has the correct spectral properties, (ii)
it is gauge invariant, (iii) it preserves causality, (iv) the
quantum kinetic equations derived with Eq.(\ref{kinetic}) coincide
with those obtained with the density matrix
technique \cite{JW84}, and (v)
it reproduces the Debye-Onsager relaxation effect \cite{MK92} .

\subsection{Quantum kinetic equation in high fields}

With the help of the gauge invariant formulation of Green's function
(\protect\ref{gaugerule}), we can write the kinetic equation
(\protect\ref{timediagonal}) in the following gauge-invariant form
\begin{eqnarray}\label{tdinv}
&&\frac{\partial}{\partial T}
f_a(k,T)+e{E}\nabla_k f_a(k,T) =\sum\limits_b\int\limits_0^{\infty} d\tau\left[  \right .
\left . \left \{ G^>_a(k-\frac{eE}{2}\tau,\tau,T-\frac{\tau}{2}),
\Sigma_{ab}^<(k-\frac{eE}{2}\tau,-\tau,T-\frac{\tau}{2}) \right \}_+ \right .\nonumber\\ 
&-&\left .
\left\{ G_a^<(k-\frac{eE}{2}\tau,\tau,T-\frac{\tau}{2}),
\Sigma_{ab}^>(k-\frac{eE}{2}\tau,-\tau,T-\frac{\tau}{2}) \right \}_+ \right]. \nonumber\\
&&
\end{eqnarray}
where all functions are now gauge-invariant. This kinetic equation is exact in time convolutions. This is necessary because gradient expansions in time are connected with linearization in electric fields and consequently
fail \protect\cite{m87}. The gradient approximation in space has been applied assuming slow varying processes in space. This local dependence of all functions on $R$ is suppressed in (\ref{tdinv}) and furtheron. We will discuss the validity in the last section (\ref{space1}).

To obtain an explicit form for the kinetic equation we have to
determine the selfenergy $\Sigma^{\grlo}$.
For the selfenergy we use the statically screened Born approximation

\begin{eqnarray}\label{self}
&&\Sigma_{ab}^{\grlo} (pR \tau T)={2 \over \hbar^2} \int \frac{d{ \bar{p}} d{ \bar{p'}} d{ p'} }
{(2\pi \hbar)^{6}} \delta({\bf p}+{\bf p'}-{\bf \bar{p}}-{\bf \bar{p'}})
V_{ab}^2({\bf p}-{\bf \bar{p}})\nonumber \\
&\times&
G_b^{\logr}(p',-\tau,R,T)G_a^{\grlo}(\bar{p},\tau,R,T)G_b^{\grlo}(\bar{p'},\tau,R,T)
\nonumber\\&&
\end{eqnarray}
with the potential given by the static Debye shape
\beq\label{pot}
V(p)={4 \pi e_a e_b \hbar^2/\epsilon_0 \over p^2 +\hbar^2 \kappa^2}
\eeq
and the static screening length $\kappa$. The latter expression is given by
\beq\label{screen}
\kappa^2=\sum\limits_c s_c {4 \pi e_c^2 n_c \over \epsilon_0 T_c}
\eeq
in equilibrium and nondegenerated limit.
Here $s_c$ is the degeneracy for specie $c$ and $T_c$ the corresponding temperature.

Introducing (\ref{self}) into (\ref{tdinv}),
the kinetic equation for two charged particle scattering with
charge $e_i$ and mass $m_i$ in high
electric fields reads
\protect\cite{MK92,JW84,Moa93}

\begin{eqnarray}\label{kinetic}
&&\frac{ \partial}{\partial T}  f_a + e {\bf E} {\pa {\bf k_a}} f_a =
\sum_b  I_{ab} \nonumber \\
\vspace*{2cm}
I_{ab} &=&
\frac{2 s_b}{\hbar^2}\int \frac{ d {\bf k'_a} d {\bf k_b} d {\bf k'_b}
}{(2\pi\hbar)^6}
\delta \left({\bf k_a}+{\bf k_b}- {\bf k'_a}-{\bf k'_b} \right)\nonumber\\
&\times&V_{ab}^2({\bf k_a}- {\bf k'_a})
\int\limits_0^{\infty} d\tau
\cos
\left\{
(\epsilon_a+\epsilon_b-{\epsilon'_a}-{\epsilon'_b}){\tau \over \hbar} \right .
- \left .
\frac{{\bf E}\tau^2}{2 \hbar}
\left(
\frac{e_a{\bf k_a}}{m_a}+ \frac{e_b{\bf k_b}}{m_b}-
\frac{e_a {\bf k'_a}}{m_a}-\frac{e_b {\bf k'_b}}{m_b}
\right)
\right\}\nonumber\\
&\times&\left\{
f'_a f'_b(1-{f}_a)(1-{f}_b)-{f}_a {f}_b (1-f'_a)(1-f'_b)
\right\}.\nonumber\\
&&
\end{eqnarray}

Here we have written, e.g., $f_b$ for $f_b(k_b-e_bE\tau,T-\tau)$
for simplicity. If we had used the conventional Kadanoff and Baym ansatz (\ref{g})  we would have obtained a factor $1/2$ in different retardations \cite{JW84}. This would lead to no relaxation effect at all \cite{MK92}.
Furthermore it is assumed, that no charge or mass transfer will occur during
the
collision. Otherwise one would obtain an additional term in the
$\cos$ function
proportional to $\tau^{3}$.
This kinetic equation is derived in second Born approximation but with complete field dependence.
Generalizations can be found for the T-matrix \cite{Mor94} approximation resulting into a field dependent Bethe-Salpeter equation or for the RPA approximation \cite{Moa93} resulting into a Lenard-Balescu kinetic equation.

Two modifications of the usual Boltzmann collision integral can be deduced
from (\ref{kinetic}):
\begin{description}
\item [(i)]{A broadening of the $\delta$-distribution function of
the energy conservation
and an additional retardation in the center-of-mass times of the
distribution functions. This is known as collisional broadening and is
a result of the finite collision duration
\protect\cite{SDKW86}. This effect can be observed even if no
external field is applied. It is interesting to remark that this
collisional broadening ensures the conservation of the total
energy \protect\cite{K75}.
If this effect is neglected one obtains the Boltzmann equation for
the field free case.}
\item [(ii)]{
The electric field modifies the broadened $\delta$- distribution
function
considerably by a term proportional to $\tau^2$. This broadening
vanishes for identical charge to mass ratios of colliding
particles. At the same time
the momentum of the distribution function becomes retarded by the
electric field. This effect is sometimes called
intra-collisional-field effect.}
\end{description}

\section{Field effects on current}

We are now interested in corrections
to the particle flux, and therefore obtain from (\ref{kinetic}) the balance equation for the momentum
\begin{equation}\label{p}
\frac{\partial}{\partial t} <{\bf k_a}>-ne_a{\bf E} =
\sum_b<{\bf k_a}I_B^{ab}>.
\end{equation}

For the Boltzmann collision integral the momentum transfer on the
right side would be zero. Here we shall find a finite value
 which will be represented as renormalization of the external field $E$
similar to the Debye-Onsager-Relaxation field in the theory of electrolyte
transport
\protect\cite{f53,FEK71,KKE66}. This effect can be shown to be a result of the
deformation of the two-particle correlation function by an applied electric
field.

To proceed we assume some important restrictions on the
distribution functions. First we assume a nondegenerate
situation, such that the Pauli blocking effects can be neglected.
In order to calculate the current for a
quasistationary plasma we choose Maxwellian distributions
\beq
f_i(p)=n_i \lambda_i^3 {\rm e}^{-{p^2 \over 2 m_i T_i}}
\eeq
with the thermal wave length $\lambda_i^2=2 \pi \hbar^2/(m_i T_i)$ and
a partial temperature for species $i$ which can be quite different
e.g. in a two - component plasma.

Then the momentum conservation in
(\ref{kinetic}) can be carried out and we get for momentum
transfer
\beq
&&<{\bf k}I_B^{ab}>=
-
\frac{2 s_b}{\hbar^2}\int \frac{d {\bf q}d {\bf Q}d {\bf
k}}{(2\pi\hbar)^9} V^2(q)
\int\limits_0^\infty d\tau
({\bf k}+e_a {\bf E} \tau) \nonumber\\
&\times&\cos\left[
 \left (-\frac{q^2}{2m_a \hbar}-\frac{{\bf k}{\bf q}}{m_a \hbar}+
\frac{{\bf q}{\bf Q}}{m_b \hbar}\right) \tau \right .+ \left . \frac{{\bf E}{\bf q}}{2 \hbar}
\left[ \frac{e_b}{m_b}-\frac{e_a}{m_a} \right] \tau^{2}
\right]
\nonumber\\
&\times&\left\{f_a({\bf k}) f_b \left({\bf Q}+\frac{\bf q}{2}\right)-
f_a({\bf k}+{\bf q}) f_b \left({\bf Q}-\frac{\bf q}{2}\right)
\right\} \nonumber\\
\eeq
where we have shifted the retardation into the distribution
functions. The second part of the distribution functions can be transformed into
the first one by ${\bf k+q } \rightarrow {\bf q}$ and ${\bf Q}-{\bf q}/2
\rightarrow {\bf Q}+{\bf q}/2$ with the result
\beq
&&<{\bf k }I_B^{ab}>=
\frac{2 s_b}{\hbar^2 (2\pi\hbar)^9}\int d {\bf
k} d {\bf q}d {\bf Q} f_b({\bf Q}) f_a({\bf k}) V^2(q) {\bf q}\nonumber\\
&\times&\int\limits_0^\infty d\tau
\cos
\left[
 \left(
 -\frac{q^2}{2 \mu \hbar}-\frac{{\bf k}{\bf q}}{m_a \hbar}+
\frac{{\bf q}{\bf Q}}{m_b \hbar}
\right)
\tau \right .+ \left .\frac{{\bf E}{\bf q}}{2 \hbar}
\left[
\frac{e_b}{m_b}-\frac{e_a}{m_a}
\right]
\tau^{2}
\right]
\eeq
with the reduced mass $\mu^{-1}=1/m_a+1/m_b$. The angular
integrations can be carried out trivially and we get
\beq\label{Ib}
&&<{\bf p}I_B^{ab}>={{\bf E} \over E} I_1\nonumber\\
&&I_1=\frac{s_b}{\hbar^{11} 4 \pi^6}\int d q q^3 V^2(q)
\int\limits_0^\infty d\tau
{\rm js}\left(
\frac{E q}{2\hbar}
\left[ \frac{e_b}{m_b}-\frac{e_a}{m_a} \right] \tau^{2}
\right) \sin({q^2 \tau \over 2 \mu \hbar}) I_2[a] I_2[b]\nonumber\\
&&
\eeq
with ${\rm js}(x)=(x \cos x-\sin x)/x^2$. The two integrals over the
distribution functions $I_2$ can be done with the result
\beq
I_2[a]&=&{\hbar m_a \over q \tau} \int\limits_0^{\infty} d k f_a(k)
\sin({k q \tau \over m_a \hbar})\nonumber\\
&=&2 \hbar^3 n_a \pi^2 {\rm e}^{-{q^2 \tau^2 T_a\over 2 \hbar^2
m_a}}
\eeq
and correspondingly $I_2[b]$.

We now introduce dimensionless variables
\beq
q&=&2\, y \, \sqrt{\mu T}\nonumber\\
t&=&{2 T \tau \over \hbar}\nonumber\\
T&=&\frac 1 2 \left ({m_b \over m_a+m_b} T_a+{m_a \over m_a+m_b}
T_b \right )\nonumber\\
e&=& {\hbar \sqrt{\mu} E \over 4 T^{3/2}}
\left[ \frac{e_b}{m_b}-\frac{e_a}{m_a} \right]
\eeq
and obtain
\beq
I_1&=&{8 n_a n_b \mu^2 T \over \pi^2 \hbar^4}
\int\limits_0^{\infty} dy y^3 V^2(2 y \sqrt{\mu T})
\int\limits_0^{\infty} dt \,{\rm js}(y t^2 e) \sin(y^2 t) {\rm e}^{-y^2
t^2}.\nonumber\\
\eeq
Using now the screened Debye potential (\ref{pot})
we finally obtain
\beq\label{integral}
I_1&=&{8 n_a n_b e_a^2 e_b^2 s_b \over T \epsilon_0^2} I_3\nonumber\\
I_3&=&\int\limits_0^{\infty} dz {z^3 \over (z^2+1)^2}
\int\limits_0^{\infty} dl \, {\rm js}(x z l^2) {\sin(z^2 l b ) \over b}
{\rm e}^{-z^2 l^2}.
\eeq
Therein we used $y=z b$ and $l=t b$ with the quantum parameter
\beq\label{b}
b^2={\hbar^2 \kappa^2 \over 4 \mu T}
\eeq
and the classical field parameter
\beq\label{x}
x=\frac e b={E \over 2 T \kappa}\left ({m_a \over m_a+m_b}
e_b-{m_b \over m_a+m_b}
e_a \right ).
\eeq
With this form (\ref{integral}) we have given an extremely usefull
representation because the field effects, contained in
x, are separated from the quantum effects, which are contained in
b. Both expansions will be performed and discussed.
To proceed we give a series expansion in the field parameter
$x$
\beq\label{sum}
I_3&=&2 \sum\limits_{k=0}^\infty (-1)^{k+1}{k+1 \over (2 k+3)!}
x^{2 k+1} I_3^{2 k+1}\nonumber\\
I_3^{2 k+1}&=&\int\limits_0^{\infty} dz {z^{2 k+4} \over (z^2+1)^2}
\int\limits_0^{\infty} dl \, l^{4 k+2} \,{\sin(z^2 l b ) \over b}
{\rm e}^{-z^2 l^2}\nonumber\\
&=&{(2 k+1)! \over 2}\int\limits_0^{\infty} dz {z^{2 -2 k} \over (z^2+1)^2}
\, {\rm _1 F_1}(2+2 k, \frac 3 2, -{b^2 z^2 \over 4}).\nonumber\\
\eeq
The last integral can be done analytically and is treated in the next section.

\subsection{Classical limit}

First we give here an exact expression for the classical limit.
We obtain from (\ref{sum}) for $b\rightarrow 0$
\beq
I_{3c}^{2 k+1}&=&{(2 k+1)! \over 2}\int\limits_0^{\infty} dz {z^{2 -2 k}
\over (z^2+1)^2}={(2 k+1)! \over 8} \pi (-1)^k (1-2 k).
\eeq
With the help of this result we can sum (\ref{sum}) with the restriction
$x<1$ from (\ref{x}) and obtain
\beq\label{classical}
I_{3c}&=&{\pi \over 8} \sum\limits_{k=0}^\infty {2 k-1 \over 2
k+3} x^{2 k+1}=-{\pi x \over 24} F(x)\nonumber\\
F(x)&=&-\frac{3}{x^2}\left [ {4-3 x^2 \over 1-x^2} -\frac 4 x {\rm
artanh}x\right ].
\eeq
This is the main result of our calculation. It gives an exact
classical result for the field dependent Debye- Onsager- relaxation- effect up
to field strengths $x<1$.

Introducing the classical result (\ref{classical}) into
(\ref{integral}) we find from (\ref{Ib}) and (\ref{p}) the following relaxation field
\begin{equation}\label{pe}
\frac{\partial}{\partial t} <{\bf k_a}>-n_ae_a{\bf E}\left (1+{\delta
E_a\over E}\right) =0
\end{equation}
with
\beq\label{form}
{\delta E_a \over E}&=&{e_a \pi \over 6 \kappa} \sum\limits_b s_b
{4 n_b e_b^2\over \epsilon_0^2 \mu}{{e_b\over m_b}-{e_a\over
m_a}\over \left ({T_b\over m_b}+{T_a\over
m_a}\right )^2} F(x).
\eeq
We see that for a plasma consisting of particles with equal charge to
mass ratios, no relaxation field appears. The link to the known
Debye- Onsager relaxation effect can be found if we assume that we
have a plasma consisting of two species with opposite charge
but equal masses and temperatures $T_a=T_b=T$. Then (\ref{form}) reduces to
\beq\label{form1}
{\delta E_a\over E}&=&-{\kappa e_a^2\over 6 \epsilon_0 T} F(x_c)
\eeq
with
\beq
x_c={e E \over T \kappa}
\eeq
and $F(x)=1+o(x^2)$ from (\ref{classical}).
This formula together with the general form (\ref{form}) is the main result of
the paper. It gives the classical relaxation effect up to any field
strength and represents a result beyond lineare response. Further it leads to a natural limitation of our treatment with respect to the field strengths which will be discussed in section (\ref{space1}). We like here to point out that the opposite special case of light particles screened by heavy ions lead to the same relaxation effect, because the limit $m_b/m_a\rightarrow \infty$ leads to (\ref{form1}) with $-x_c$. Because of the even character of $F(x)$ this leads to the same expression as in the case of identical oppositely charged particles.

\subsection{Quantum correction}
The complete quantum case of (\protect\ref{sum}) can be given by performing the integral. The result, which can be verified by MATHEMATICA reads
\beq\label{quantum}
&&I_3^{2 k+1}={1\over 4^{k+2}}\,{b^{1 + 2\,k}}\,{\sqrt{\pi }}\,
      {{\rm \Gamma}(-{1\over 2} - k)\,
      {\rm \Gamma}({5\over 2} + 3\,k)\, \over (k+1)!}
      {\rm _2F_2}(\{ 2,{5\over 2} + 3\,k\} ,
       \{ {3\over 2} + k,2 + k\} ,{{{b^2}}\over 4})
     \nonumber\\
  &+& ( -1 )^k\,\pi \,
      (2 k+1)!\,
      \left( 3\,{\rm _1F_1}(2 + 2\,k,
          {3\over 2},{{{b^2}}\over 4}) \right. + \left . 
        2\,{b^2}\,(1+k)\,{\rm _1F_1}(3 + 2\,k,
          {5\over 2},{{{b^2}}\over 4}) \right)
\nonumber\\
&&
\eeq

With the help of this integral (\ref{quantum}) we have given the complete quantum result of the series expansion with respect to electric fields (\ref{sum}). 
It is easy to create any kind of degree of response by this way. Only odd exponents of the fields appear in (\ref{sum}) which gives an even series for $F(x)$ in (\ref{classical}). In the following we give only the first two parts of the expansion with respect to the field. The quantum effects are included completely.

\subsubsection{Quantum lineare response}
The first quantum correction reads

\beq
&&I_3^1(k=0)={\pi \over 8}  \,\left( 1 + {b^2\over 4} - 
       \frac b 2 \,{\sqrt{\pi }} {{\rm e}^{{{{b^2}}\over 4}}} \left ( \frac 3 2 +{b^2 \over 4} \right )
        {\rm erfc}({b\over 2})  \right).
\nonumber\\
&&
\eeq
Renaming the quantum parameter $b/2$ from (\ref{b}) into $x$, the result of \cite{MK92} is reproduced by a different way of calculation. Unfortunately in \cite{MK92} a misprint occured where a factor $b/2$ in front of the $erfc$ was missing.

\subsubsection{Cubic response quantum result}

As a next term we can give the cubic Debye- Onsager field effect in complete quantum form
\beq
&&I_3^3(k=1)={\pi \over 64 b_1}  \,\left( 15 \sqrt{\pi} + 18 b_1 - 118 b_1^3-72 b_1^5-8 b_1^7 \right .\nonumber\\
&+&\left .{\sqrt{\pi }} {\rm e}^{b_1^2} \left ( -15+15 b_1^2+150 b_1^4+76 b_1^6+8 b_1^8 \right )
        {\rm erfc}(b_1)  \right)
\nonumber\\
&&\eeq
with $b_1=b/2$.

\begin{figure}
\epsfxsize=15cm
\centerline{\epsffile{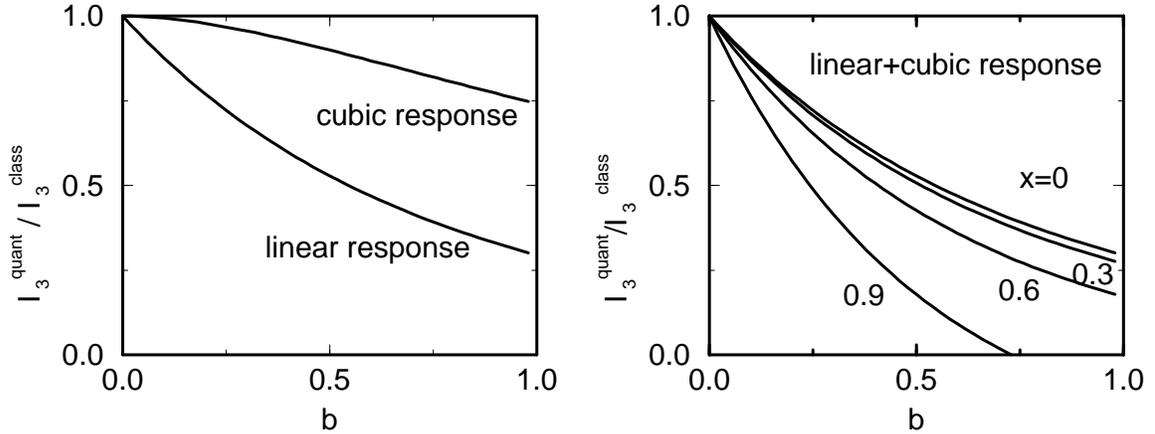}}
\vspace{2ex}
\caption{\label{2}The ratio of quantum to classical 
Debye-Onsager relaxation effect (\protect\ref{form}) versus quantum parameter $b$ of (\protect\ref{b}). In the left hand figure the lineare and cubic response result, $I_3^1$ and $I_3^3$ of (\protect\ref{sum}) is plotted separately. In the right hand figure we give the relaxation effect up to cubic terms in field parameter $x$ of (\protect\ref{x}) for different field strength.}
\end{figure}

In the Fig. \ref{2}a we plot the quantum versus classical result for 
linear response and cubic response versus the quantum parameter (\ref{b}). We see that the cubic response is less influenced by quantum effects than the lineare response result. The general observation is that the quantum effects lower the classical result for the relaxation effect. A detailed analysis of quantum effects on the lineare response can be found in \cite{MK92}. 

\vspace{2ex}

InFig. \ref{2}b we give the ratio of quantum to classical result for the relaxation effect up to cubic terms in fields for different field strengths represented by values of x from (\ref{x}). We see that the quantum effects become more important with increasing field strength. The effect of sign change for field strengths $x>0.7$ described above can also be seen in the quantum effects at certain values of $b$.

\subsection{Validity of the Calculation \label{space1}}
We have seen that the electric field is limited to values $x<<1$ for $x$ from (\ref{x}).
For equal masses and temperatures of plasma components this condition translates into
\beq\label{cond1}
E<{\kappa T\over e}.
\eeq
In this limit F(x) has a singularity and no quasi equilibrated transport is possible, i.e. no thermal distributions are pertained in the system. Then we have to take into account nonthermal field dependent distributions which have been employed to study nonlinear conductivity \cite{MSK93,KMSR92,MK93}.

The condition (\ref{cond1}) allows for different physical interpretations. Within the picture of the screening cloud we can rewrite (\ref{cond1}) into
\beq
e E < m {v_{\rm th}^2 \over r_D}.
\eeq
This means that a particle moving on the radius of the screening cloud $r_D=1/\kappa$ with thermal velocity $v_{\rm th}^2=T/m$ should not be pulled away by the acting field force. This can be considered as a condition for the possibility of a screening cloud at all in high fields. For fields higher than this limit we will not have screening at all.
We can discuss this limit also via the energy density which can be reached in a system by the applied field. We can reformulate once more the condition (\ref{cond1}) to find equivalently
\beq
{\epsilon_0 E^2 \over 4 \pi}<n T.
\eeq
This means that we have essentially nonthermal effects to be expected
if the energy density of the field becomes comparable with the thermal energy density.

The validity criterion (\ref{cond1}) can now be used to check the weak space inhomogeneity which has been assumed during our calculation. Quasi- equilibrium in a plasma system with fields can only be assumed if the field current is accompanied by an equivalent diffusion current
\beq
j_{\rm field}=e \mu E n=-j_{\rm diff}=eD{d n \over dx}
\eeq
using the Einstein  condition $\mu=eD/T$ one gets
\beq
eE=T \frac 1 n {d n \over dx}.
\eeq
Combining this elementary consideration with our condition (\ref{cond1})
we obtain a limitation for space gradients
\beq\label{space}
{d n \over d\,(\kappa \,x)}<n
\eeq
where our treatment of field effects and kinetic equation is applicable.

\section{Discussion}

\begin{figure}
\epsfxsize=9cm
\centerline{\epsffile{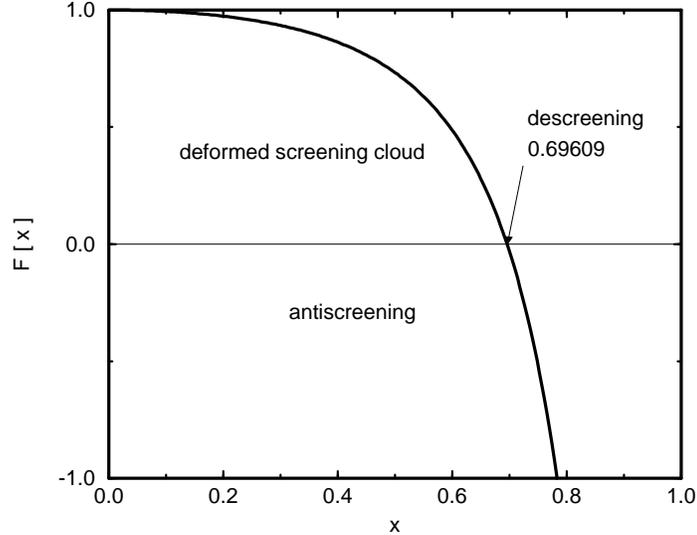}}
\vspace{2ex}
\caption{\label{1}The scaled nonlinear Debye-Onsager relaxation effect (\protect\ref{form}) or (\protect\ref{form1}) versus the field parameter (\protect\ref{x}). The lineare response regime is given by small numbers of $x$ where no remarkable deviation from classical Debye result (\protect\ref{debye}) is seen. The screening cloud is deformed further with increasing fields until no relaxation occurs at $x\approx 0.7$. This is complete descreening. Further increase of fields change the sign of the relaxation effect and the moving particles become accelerated instead of decelerated. This range is indicated by antiscreening.}
\end{figure}

It is interesting to discuss the classical nonlinear relaxation effect via the function $F(x)$ in (\ref{classical}). In figure \ref{1} we have plotted the corresponding dependence. One recognizes that the linear response regime is possible until $x\approx 0.2$. Then the relaxation effect becomes lower with increasing fields until it leads to a complete vanishing at $x=0.69609$. At this point no screening cloud is established and therefore no relaxation effect present. If we increase the field we see that the effect changes the sign. This range means that a charge is accelerated more than the applied field would allow for because of the surrounding opposite charges. We call this region antiscreening.
It occurs in the range
\beq
{\kappa T\over e}>E> 0.7 {\kappa T\over e}.
\eeq
Antiscreening means that the field is already blowing away the screening cloud. However, the charged particle is feeling an accumulation of opposite charges in front of its path. This can be considered as an analogue to the Doppler effect. We will call it Charge - Doppler effect (CD- effect).
This has consequences on the conductivity which has to increase within this range. Of course, the total energy of the system is not changed, because this acceleration comes from thermal energy of other particles.

\vspace{2ex}

The interpretation of accumulated charges in front of a moving charge is underlined by consideration of a very heavy projectile with mass $m_a$ which goes through a plasma of lighter particles $m_b$. Then we get from (\ref{form}) also a sign change. We can imagine this effect as the accumulation of plasma charges in front of the moving heavy ion. This, of course, gives a field additional to the external field.
As an application of this nonlinear Debye Onsager relaxation effect we may think of stopping power of a heavy ion in a plasma system \cite{MR96}. Within the discussed special field range of the CD-effect we expect a slight acceleration of ions and reduction of the stopping power. This is especially of importance for current experiments, where heavy ion beams are stopped in a nonlinear plasma as target.

\acknowledgements

Helpful discussions with H.P. Pavel and S. Schmidt are grateful acknowledged.
This work was supported from the BMBF (Germany) under contract
Nr. 06R0745(0).

%\bibliography{kmsr,kmsr1,kmsr2,kmsr3,kmsr4,kmsr5,kmsr6,kmsr7,vac1,spin,delay}
%\bibliographystyle{prsty}

\end{document}